\documentclass[10pt,technote]{IEEEtran}
\usepackage{amsmath,amsfonts}
\usepackage{algorithmic}
\usepackage{algorithm}
\usepackage{array}
\usepackage[caption=false,font=normalsize,labelfont=sf,textfont=sf]{subfig}
\usepackage{textcomp}
\usepackage{acronym}
\usepackage{stfloats}
\usepackage{url}
\usepackage{verbatim}
\usepackage{graphicx}
\usepackage{cite}

\usepackage{amsmath,amsfonts,amssymb}
\usepackage{algorithmic}
\usepackage{array}
\usepackage{tikz}
\usepackage[caption=false,font=normalsize,labelfont=sf,textfont=sf]{subfig}
\usepackage{textcomp}
\usepackage{lettrine}
\usepackage[none]{hyphenat}
\usepackage{placeins}
\usetikzlibrary{positioning}
\usepackage{array,tabularx}
\usepackage{textcomp}
\usepackage{cite}
\usepackage{stfloats}
\usepackage{bm}
\usepackage{graphics}
\usepackage{float}
\usepackage[center]{caption}
\usepackage{url}
\usepackage[font=small,labelfont=bf,
   justification=justified,
   format=plain]{caption}
\usepackage{verbatim}
\usepackage{graphicx}
\usepackage{acronym}
\def\BibTeX{{\rm B\kern-.05em{\sc i\kern-.025em b}\kern-.08em
    T\kern-.1667em\lower.7ex\hbox{E}\kern-.125emX}}
\usepackage{balance}

\newcolumntype{C}[1]{>{\centering\let\newline\\\arraybackslash\hspace{0pt}}m{#1}}

\acrodef{5G}[5G]{fifth generation}
\acrodef{HAP}[HAP]{high-altitude platform}
\acrodef{LTE}[LTE]{long-term evolution}
\acrodef{OTFS}[OTFS]{orthogonal time frequency space}
\acrodef{BER}[BER]{bit error rate}
\acrodef{MIMO}[MIMO]{multiple input multiple output}
\acrodef{OFDM}[OFDM]{orthogonal frequency division multiplexing}
\acrodef{LEO}[LEO]{low earth orbit}
\acrodef{NSB}[NSB]{null steering beamforming}
\acrodef{DD}[DD]{delay-doppler}
\acrodef{TF}[TF]{time-frequency}
\acrodef{QAM}[QAM]{quadrature amplitude modulation}
\acrodef{ISFFT}[ISFFT]{inverse symplectic finite Fourier transform}
\acrodef{SFFT}[SFFT]{symplectic finite Fourier transform}
\acrodef{AWGN}[AWGN]{additive white Gaussian noise}
\acrodef{LoS}[LoS]{line-of-sight}
\acrodef{NLoS}[NLoS]{non line-of-sight}

\begin{document}

\title{Integrating OTFS in Airplane-Aided Next-Generation Networking}

\author{Ashok S Kumar,  Shashank Shekhar,  Gokularam Muthukrishnan, Muralikrishnan Srinivasan, \IEEEmembership{Member, IEEE}, \\Sheetal Kalyani, \IEEEmembership{Senior Member, IEEE}

\thanks{A. S Kumar, and S. Kalyani are with the Department of Electrical Engineering, Indian Institute of Technology Madras, Chennai, India (e-mail: \{ee22d023@smail, skalyani@ee\}.iitm.ac.in).}
\thanks{Shashank Shekhar is with the Department of Electronics and Communication Engineering, PSG iTech, Coimbatore, India (e-mail: shashank@psgitech.ac.in).}
\thanks{Gokularam Muthukrishnan is with the India Urban Data Exchange, Center of Data for Public Good,
Indian Institute of Science, Bengaluru, India. (e-mail: gokularam.m@datakaveri.org).}
\thanks{Muralikrishnan Srinivasan is with the Department of Electronics Engineering, Indian Institute of Technology (BHU), Varanasi, India (e-mail: muralikrishnan.ece@iitbhu.ac.in).}

}



\maketitle

\begin{abstract}
Next-generation networks explore the opportunistic assistance of airliner/high-altitude platforms (HAPs) in delivering high data rates for terrestrial networks to ensure consistent and reliable communication. When an airliner/HAP moves at very high speeds, its mobility has a substantial impact on ensuring seamless connectivity, stable signal strength, and reliable data transmission. Orthogonal time frequency space (OTFS) modulation has been shown to provide notable improvement in performance when handling Doppler effects during high-mobility situations. This paper presents an OTFS-based airplane-aided next-generation networking system. In the proposed system, the airliner/HAPs are equipped with a planar antenna array that applies null steering beamforming (NSB) at the transmitter for communication with terrestrial users. A comprehensive performance comparison between OTFS and orthogonal frequency division multiplexing (OFDM) is performed under varying airliner altitude, velocity, array dimension, and Rician factor conditions. The simulation results show that OTFS consistently outperforms OFDM, achieving a lower bit error rate (BER) and more stable performance across different airliner altitudes, velocities, array dimensions, and propagation environments.

\end{abstract}


\section{Introduction}

\lettrine{N}{ext-generation} wireless standards, including sixth generation (6G), are expected to support dynamic channel conditions, particularly in high-mobility scenarios, while managing increased traffic demands and enabling new applications, especially in rural areas. Combining terrestrial and aerial networks can significantly improve data rates in these regions \cite{8760401}. Aerial platforms, such as airliner/\ac{HAP}, offer the potential to strengthen cellular communication in underserved areas\cite{8760401,khawaja2019survey}. However, most integration attempts are based on the congested \ac{LTE} bands, highlighting the need for innovative architectures that merge terrestrial and space networks. Analysis and simulations in work \cite{9491998} indicate that airplane-aided integrated networks can provide high-capacity solutions to address coverage gaps in next-generation wireless systems. However, such aerial platforms often operate in high-mobility scenarios, where Doppler effects severely degrade communication performance.  

\Ac{OTFS} modulation has emerged as a promising candidate to mitigate Doppler effects in high-mobility environments \cite{7925924,mohammed2021derivation}. \ac{OTFS} modulation is a two-dimensional modulation technique that processes data symbols in the \ac{DD} domain \cite{wei2021orthogonal}. The channel representation in the \ac{DD} domain is typically sparse, allowing for more effective channel estimation and equalization \cite{9661102}. In \cite{ramachandran2020otfs}, the authors investigated signal detection and channel estimation in \ac{MIMO}-based \ac{OTFS} systems and found that \ac{OTFS} significantly outperforms \ac{MIMO}-\ac{OFDM} in high-speed environments, specifically at 4 GHz and 28 GHz. The simulation results in \cite{ ramachandran2020otfs, 9404861} demonstrate that \ac{OTFS} modulation outperforms conventional \ac{OFDM} in high-mobility conditions. Furthermore, \cite{9904495} explored \ac{OTFS} modulation in multipath channels where each signal path experiences distinct rapid fading. 

In this paper, we propose an \ac{OTFS}-based airplane-aided integrated next-generation networking system, designed to address the challenges of high mobility environments. Our study focuses on the impact of increased mobility on the integration of airliner/\ac{HAP}s with planar antenna arrays for seamless communication with terrestrial users and base stations via \ac{OTFS}. We employ \ac{NSB} beamforming technique, which ensures efficient and focused signal delivery to intended users and mitigates interference. At the receiver, zero-forcing (ZF) equalization is applied in the DD domain for OTFS \cite{singh2022ber}. The novelty of this work lies in demonstrating the feasibility of OTFS modulation in airliner/HAP-aided high-mobility networking, where long propagation distances, severe Doppler shifts, and large antenna arrays fundamentally alter the system behavior. Although prior work \cite{sinha2020otfs, mohammed2021time, nordio2025joint, bora2022irs} has investigated OTFS in various high-mobility scenarios, these studies have largely considered terrestrial or UAV-centric settings with limited coverage range. However, its potential for long-range, high-altitude airborne networking, particularly in the presence of multi-tier interference and large-scale antenna arrays, has not been explored. Our objective is to investigate how OTFS, when integrated with NSB beamforming and DD domain equalization, performs against OFDM in terms of \ac{BER}-signal to noise ratio (SNR) characteristics under varying Rician factors, mobility levels, altitudes, and antenna dimensions, thereby demonstrating its suitability for next-generation airborne communication systems.


The notation used in the paper is as follows: transpose and conjugate transpose of a vector or matrix are denoted by $(.)^T$ and $(.)^H$, respectively, $\|.\|$ represents the Euclidean norm. Scalars are represented in lowercase, while vectors and matrices are denoted by lowercase bold and uppercase bold, respectively.

\section{System Model}
\label{Sec:System model}
We consider a system design, in which an airliner/\ac{HAP} is positioned at an altitude $H_{t}$ above the center of a macro-cell with radius $R$, where $R \gg 5$ km, while maintaining an inter-airliner separation of at least $10$ km \cite{9491998}. This macro-cell is divided into micro-cells of radius $r$, each supporting a single time-frequency block, where $r \ll R$. The intended user, either cellular or \ac{LTE} base stations, can be located at $(x_{0}, y_{0}, -H_{t})$, representing the micro-cell of interest (MCI). \begin{figure}[!ht]
\includegraphics[width=3.3 in]{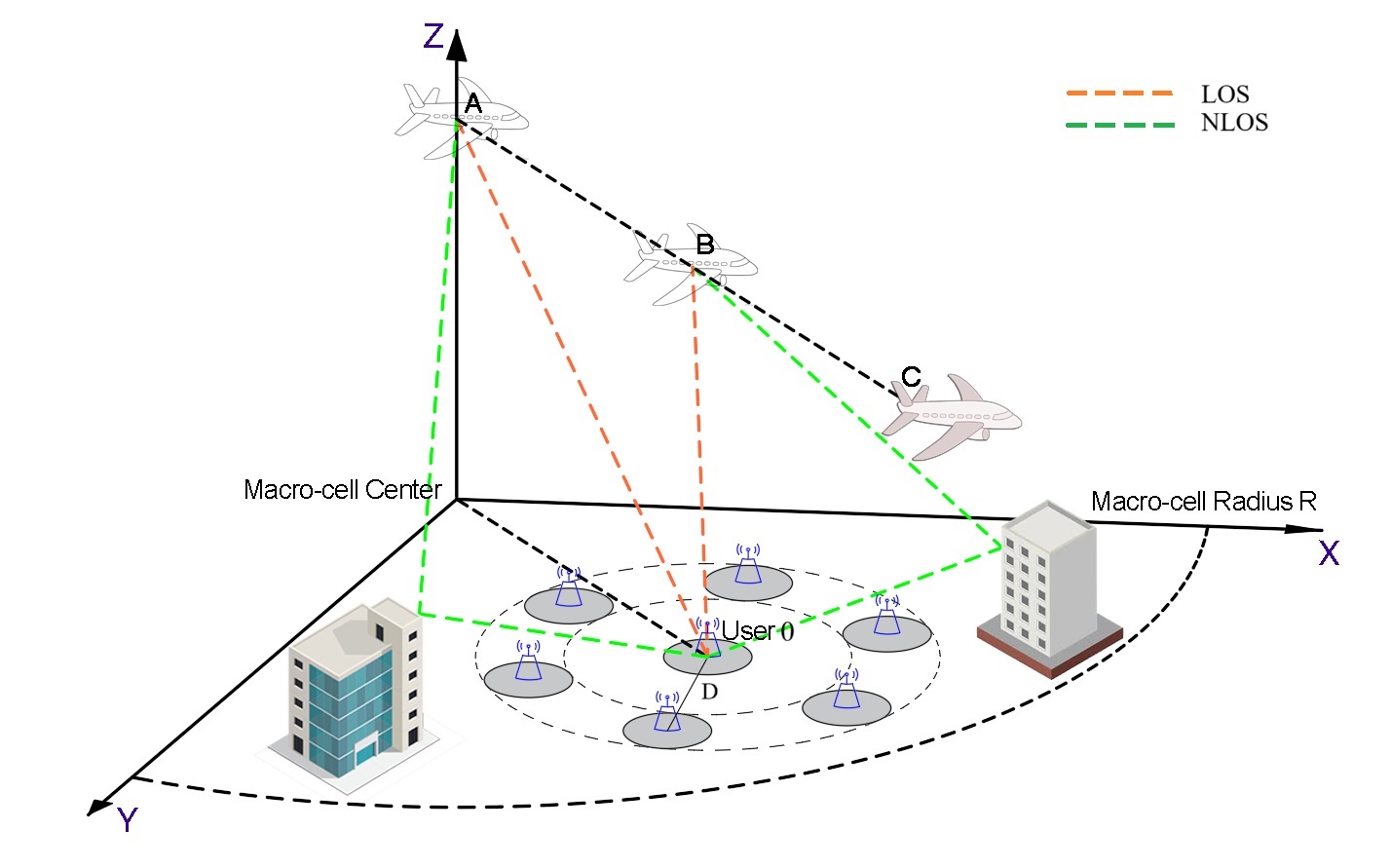}
\centering
    \caption{The airliner/HAP is initially assumed to be located at the point A $(0, 0, 0)$. The Macro-cell radius $R$ denotes the distance measured from the center of the macro-cell. At point A, there is an initial \ac{LoS} and \ac{NLoS} condition between the user and the airliner. As the airliner/HAP moves from point A to point C, the \ac{LoS} and \ac{NLoS} conditions may change over time.}
\label{fig220}

\end{figure}

The interference comes from $Q$ tiers of micro-cells surrounding the MCI. The tier $1$ contains $B_{1}$ users located at a distance $D$
from the MCI, where $D$ represents the frequency reuse distance. Similarly, tier $2$ contains $B_{2}$ users at a distance $2D$, and so on, until tier $Q$ contains $B_{Q}$ users at a distance of $QD$ from the MCI. The total interference is from $B_{t}$ users where $B_{t} = B_{1} + B_{2} + \ldots + B_{Q}$, with the number of tiers selected based on the user's proximity to the macro-cell center\footnote{For example, $3$ tiers are sufficient near the center, while $5$ tiers may be required at the edge.}. Each user is assumed to have a single antenna, with coordinates $(x_{i}, y_{i}, -H_{t})$ for $i=0, 1, 2, \ldots, B_{t}$. The airliner has a uniform planar antenna array of $L \times L$ elements, spaced $\frac{\lambda}{2}$ apart, where $\lambda$ is the carrier wavelength. The $(j,k)^{th}$ element’s position is given by $\left(x_{j}, y_{k}, 0\right)$, where $x_{j} = \left[-\frac{L-1}{2} + (j-1)\right] \frac{\lambda}{2}$ and $y_{k} = \left[-\frac{L-1}{2} + (k-1)\right] \frac{\lambda}{2}$. Let an $L^{2}\times1$  vector $\bm{\alpha}_{i}$, represent the steering vector of the $i^{th}$ user for all elements of the planar array. The value of the steering vector corresponding to the location of the user relative to the $(j,k)^{th}$ element in the uniform planar array is expressed as $\exp \left[j \frac{2 \pi}{\lambda}\left(x_{j} \psi_{i}^{x}+y_{k} \psi_{i}^{y}\right)\right]$, where $\psi_{i}^{x} = \sin \theta_{i}^{z} \cos \theta_{i}^{a}$ and $\psi_{i}^{y} = \sin \theta_{i}^{z} \sin \theta_{i}^{a}$. The angles $\theta_{i}^{z}$ (zenith) and $\theta_{i}^{a}$ (azimuth) for the $i^{th}$ user are given by:
\[
\theta_{i}^{z} = \tan^{-1}\left(\frac{\sqrt{x_{i}^{2} + y_{i}^{2}}}{-H_{t}}\right), \quad \theta_{i}^{a} = \tan^{-1}\left(\frac{y_{i}}{x_{i}}\right),
\]
where $i=0$ denotes the intended user, and $i=1, 2, \ldots, B_{t}$ represents interfering users. Fig. \ref{fig220} shows the overall architecture of the system. A tier of users that interfere $(B_{1}=6)$ with a reuse distance of $D$ is shown. 

\begin{figure*}[!ht]
\centering
\includegraphics[width=4in]{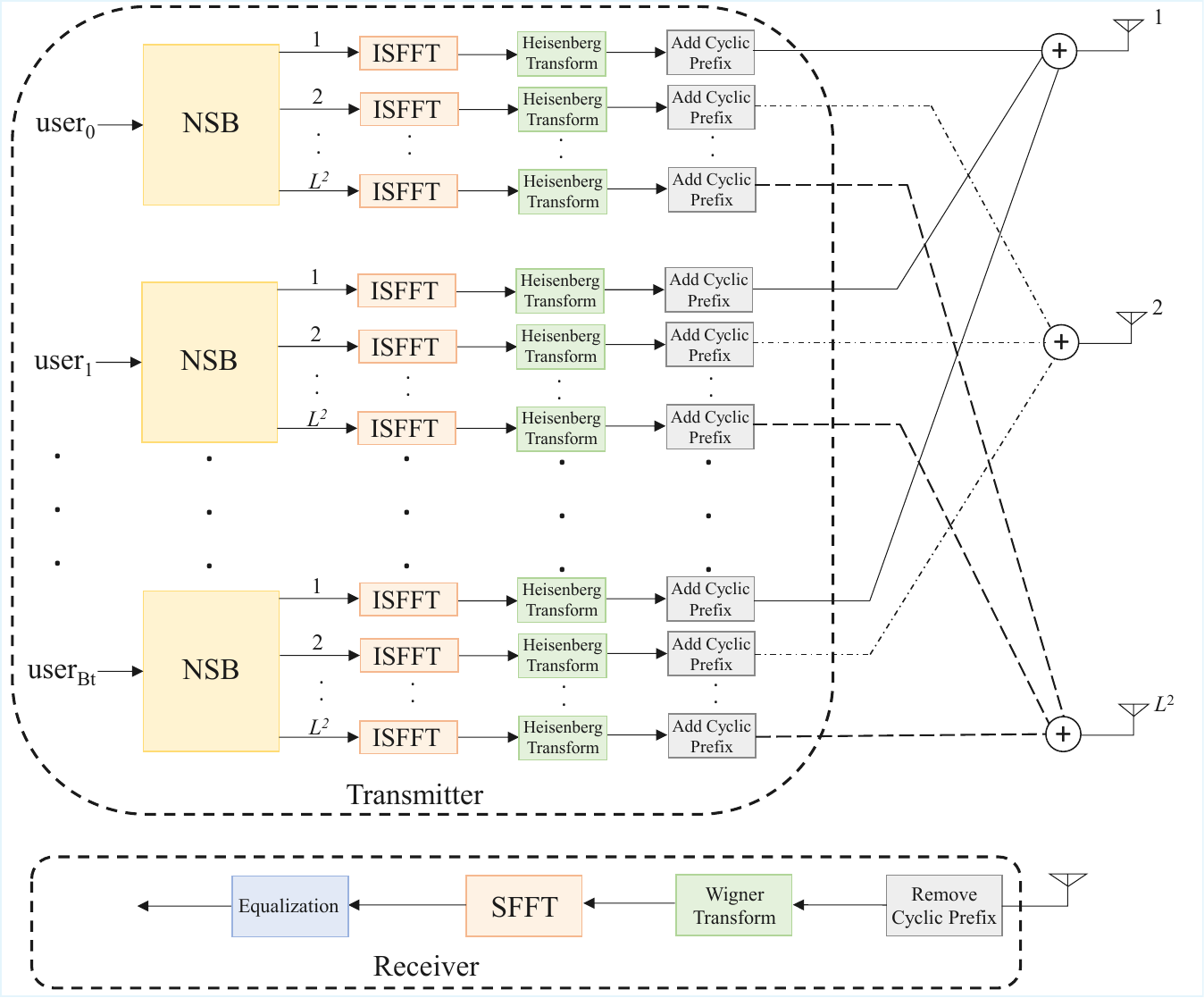}
\caption{Block diagram of the proposed OTFS system for airplane-aided integrated networking}
\label{fig221}
\end{figure*}

The airliner/HAP transmits OTFS frames to users on the ground. OTFS, being a robust modulation scheme for time-varying channels, offers resilience to DD shifts caused by airliner/HAP movement. This is especially crucial when the airliner/HAP is moving at high speeds, introducing significant Doppler shifts that can degrade conventional modulation schemes. The proposed \ac{OTFS} system for airplane-aided networking is illustrated in Fig. \ref{fig221}. To enhance spectral efficiency and mitigate interference, we employ \ac{NSB}, similar to ZF precoding \cite{yoo2003multipath}. Let $\tilde{\mathbf{\alpha}}_{i}(j)$ represent \ac{NSB} coefficient on the $j^{th}$ antenna element for the $i^{th}$ user. \ac{NSB} vector of the $i^{th}$ user is represented by 
$\tilde{\bm{\alpha}}_i=\bm{\alpha}_i-\mathbf{\Lambda}_i\left(\mathbf{\Lambda}_i^H \mathbf{\Lambda}_i\right)^{-1} \mathbf{\Lambda}_i^H \bm{\alpha}_i, \forall i=0,1, \ldots, B_t $, where $\mathbf{\Lambda}_i=\left[\bm{\alpha}_0 \,\,\bm{\alpha}_1 \ldots \bm{\alpha}_{i-1} \,\,\bm{\alpha}_{i+1} \ldots \bm{\alpha}_{B_{t}}\right]$ constitutes the matrix having its columns as the steering vectors, except for $\bm{\alpha}_i$.
 \ac{NSB} is achieved by solving an optimization problem to minimize $\|\tilde{\bm{\alpha}}_i - \bm{\alpha}_i\|^2$, subject to $\tilde{\bm{\alpha}}_i^H \mathbf{\Lambda}_i = 0$. This ensures maximum signal power toward the desired user while nullifying the response towards interfering users. Intuitively, NSB projects the desired steering vector onto the subspace orthogonal to all interference directions, preserving the main beam while nulling unwanted directions. We now describe the \ac{OTFS} frame structure used for data transmission.

Each \ac{OTFS} frame has $M$ sub-carriers and $N$ time slots. The information bits of the user $i$ are mapped to a quadrature
amplitude modulation (4-QAM)  symbol set $A_{\mathrm{DD}}^i[k, l]$ in the \ac{DD} domain, where $k = 0, \ldots, N-1$ and $l = 0, \ldots, M-1$. The \ac{DD} domain symbol is transformed to the \ac{TF} domain using the \ac{ISFFT}:
\begin{equation}
    A_{\mathrm{TF}}^i[n, m] = \frac{1}{\sqrt{NM}} \sum_{k=0}^{N-1} \sum_{l=0}^{M-1} A_{\mathrm{DD}}^i[k, l] e^{j 2\pi\left(\frac{nk}{N} - \frac{ml}{M}\right)}.
\end{equation}
The \ac{TF} domain signal $A_{\mathrm{TF}}^i[n,m]$ spans a lattice $L=\{(nT, m\Delta f)\}$, where $\Delta f$ is the sub-carrier spacing and $T = 1/\Delta f$. The transmitted signal $s_i(t)$ in the time domain is given by:
\begin{equation}
   s_i(t) = \sum_{n=0}^{N-1} \sum_{m=0}^{M-1} A_{\mathrm{TF}}^i[n,m] g_{\text{tx}}(t-nT) e^{j 2 \pi m \Delta f (t-nT)}, 
\end{equation}
where $g_{\text{tx}}(t)$ represents the ideal rectangular
pulse-shaping filter. The transmit signal from the $j^{th}$ antenna element, $\tilde{s}_j(t)$, is constructed as a weighted sum of the signals for all users, as follows:
\begin{equation}
    \tilde{s}_j(t) = \sum_{i=0}^{B_{t}} \tilde{\alpha}_i(j)\, s_i(t).
\end{equation}
This transmitted signal is then subject to channel effects, including significant Doppler shifts introduced by high-speed movements, before it reaches the user's receiver. In the next section, we examine the reception process on the user side.

\section{OTFS Reception at the User}
\label{OTFS reception}

The high-speed movement of the airliner/\ac{HAP} introduces significant Doppler shifts. The Doppler shift is assumed to be constant for different paths of the same user due to stationary scatterers during the observed time. In this setup, we consider an \ac{LoS} and an \ac{NLoS} path, with a maximum delay spread of $150$ ns, as specified in the 3GPP D5 deployment scenario \cite{3GPPTR37885}. 
The time-varying channel impulse response between the transmitter and the $j^{th}$ antenna element, denoted as $h_{0,j}(t, \tau)$. The \ac{LoS} and \ac{NLoS} terms are weighted by the Rician $K$ factor ($\kappa_{d}$), path gains, Doppler shifts, and respective delays:

{\small
\begin{align}
h_{0,j}(t,\tau) =\; & \sqrt{\frac{\kappa_{d}}{\kappa_{d}+1}}\, h_{0,j}^{\mathrm{LoS}}(t)\, \alpha_{0}(j) \,
e^{j 2 \pi \nu_{0,j}^{\mathrm{LoS}} (t - \tau_{0,j}^{\mathrm{LoS}})} \delta(\tau - \tau_{0,j}^{\mathrm{LoS}}) \notag \\
&\hspace{-1em} + \sqrt{\frac{1}{\kappa_{d}+1}}\, h_{0,j}^{\mathrm{NLoS}}(t) \,
e^{j 2 \pi \nu_{0,j}^{\mathrm{NLoS}}(t-\tau_{0,j}^{\mathrm{NLoS}})} \delta(\tau-\tau_{0,j}^{\mathrm{NLoS}}).
\end{align}
}
Here, $h_{0,j}^{\mathrm{LoS}}(t)$ is the \ac{LoS} gain, assumed to be unity, while the \ac{NLoS} component $h_{0,j}^{\mathrm{NLoS}}(t)$ is modeled as a complex Gaussian distribution $\mathcal{CN}(0,1)$. The delay $\tau_{0,j}(t)$ between the airliner and the user $0$ at time $t$ is calculated as $\tau_{0,j}(t)=\frac{d_{0}(t)}{c}$, where $d_{0}(t) = \|D_{0} - D_{\mathrm{HAP}(t)}\|$ is the distance between the user $0$ and the airliner at time $t$, with $D_{0}$ and $D_{\mathrm{HAP}(t)}$ representing the position vectors of user $0$ and the airliner, respectively. Finally, the Doppler frequency shift $\nu_{0,j}(t)$ is expressed as:
\begin{equation}
    \nu_{0,j}(t) = \frac{f_{c}v(t)\cos(\theta_{0}(t))}{c},
\end{equation}
where $\theta_{0}(t)$ is the angle between the airliner and the user $0$ at time $t$, $c$ is the speed of light, $f_{c}$ is the carrier frequency, and $v(t)$ is the velocity of the airliner. The time domain representation of the received signal for the user $0$
is:
\begin{equation}
\begin{aligned}
r_{0}(t)
&= \sqrt{P_{r}} \sum_{j=1}^{L^2} \int h_{0,j}(t,\tau)\,
   \tilde{\mathbf{\alpha}}_{0}(j)\, s_{0}(t-\tau)\, d\tau \\
&\quad + \sqrt{P_{r}} \sum_{j=1}^{L^2} \sum_{i=1}^{B_{t}}
   \int h_{0,j}(t,\tau)\,
   \tilde{\mathbf{\alpha}}_{i}(j)\, s_{i}(t-\tau)\, d\tau + w(t) \\
&= \sqrt{P_{r}} \sum_{j=1}^{L^2} \int h_{0,j}(t,\tau)\,
   \tilde{s}_{j}(t-\tau)\, d\tau + w(t).
\end{aligned}
\end{equation}
In the above formulation, $i=0$ corresponds to the intended (desired) user, while $i=1,2,\ldots, B_t$ represent the interfering users. Thus, the received signal expression in Eq.~(6) inherently captures both the desired and multiuser interference components. The final compact form is obtained by combining all user transmissions into the aggregate radiated signal $\tilde{s}_j(t)$.
Here, $P_{r}$ is the power received, which depends on multiple factors such as the transmitted power from the airliner, antenna gains, atmospheric losses, back-off loss, and other transmission losses. Here, $w(t)$ denotes the complex additive white Gaussian noise with variance $\sigma^2$.
$\sigma^{2}$ depends on Boltzmann's constant $k$, temperature $T$ (in Kelvins), bandwidth $B$ and the receiver's noise figure, $N_{f}$. The noise power is given by $\sigma^{2}=kTBN_{f}$. At the \ac{OTFS} receiver, the received time-domain signal $r_{0}(t)$ is transformed into the \ac{TF} domain using the Wigner transform \cite{7925924,wei2021orthogonal}:
\begin{equation}
    B_{\mathrm{TF}}[n, m]=\int_{-\infty}^{\infty} r_{0}(t) g_{\text{rx}}^{*}(t-n T) e^{-j 2 \pi m \Delta f(t-n T)} d t,
\end{equation}

where $g_{\text{rx}}$ is the receiver ideal rectangular
pulse-shaping filter. The \ac{TF} domain signal $B_{\mathrm{TF}}[n, m]$ is further transformed into the \ac{DD} domain using the \ac{SFFT} \cite{7925924,wei2021orthogonal}:

\begin{equation}
    B_{\mathrm{DD}}[k, l]=\frac{1}{\sqrt{N M}} \sum_{n=0}^{N-1} \sum_{m=0}^{M-1} B_{\mathrm{TF}}[n, m] e^{-j 2 \pi\left(\frac{n k}{N}-\frac{m l}{N}\right)}.
\end{equation}
After converting the signal to the DD domain, equalization is carried out in the DD domain.

\section{Simulation Results}
\label{Simulation}
We now illustrate the numerical simulations to assess how OTFS and OFDM perform under varying propagation and system conditions, particularly in high-Doppler environments. The OFDM is the baseline waveform in current wireless standards and provides a fair benchmark to highlight the Doppler resilience of OTFS. The simulation parameters used in this study are summarized in Table \ref{tab:allmethods}. We use a $28$ GHz carrier, a widely adopted mmWave band for 5G/6G. For simplicity, we assume that all users (both intended and interfering) remain static throughout the \ac{HAP}s journey, with stationary scatterers as well. The analysis begins with the airliner/HAP positioned at the known coordinates $(0, 0, 0)$ at the initial time, $t = 0$. The intended user is uniformly distributed within the MCI at coordinates $(x_{0}, y_{0}, -H_{t})$. Similarly, interference users are uniformly distributed within the MCI at the coordinates $(x_{i}, y_{i}, -H_{t})$ for $i = 1, 2, \ldots, B_{t}$. For a frequency reuse factor of $7$, the reuse distance
is fixed at approximately $D = 4r$. We adopt a five-tier model for interfering users, with their centers set to $4\tilde{k}r$, where $\tilde{k} = 1, 2, \ldots, 5$ from the MCI \cite{9491998}. Given this initial setup, the positions of both the intended and interfering users are clearly defined, allowing for precise calculation of the azimuth and zenith angles, as well as the distances between the users and the airliner/HAP. The Rician fading factor, $\kappa_{d}$, is fixed at $10$ dB to model the \ac{LoS} component. The proposed system is capable of supporting a maximum velocity given by $V_{\text{max}} = \frac{c \Delta f}{2 f_{c}}$, where $\Delta f$ is the sub-carrier spacing\footnote{For a carrier frequency of $28$ GHz and a speed of $150$ m/s, this results in a maximum Doppler shift of $\pm 14$ kHz. This highlights the significant time-selectivity of the channel, especially at higher velocities}. For a fair comparison with \ac{OTFS}, we use an \ac{OFDM} system with $512$ subcarriers that maintains the same bandwidth and time duration, with a cyclic prefix length of $4$ samples for each OFDM symbol. We perform ZF equalization in the DD domain for OTFS and in the frequency domain for OFDM.

\begin{figure}[h]

\includegraphics[width=2.8 in,height=2.05in]{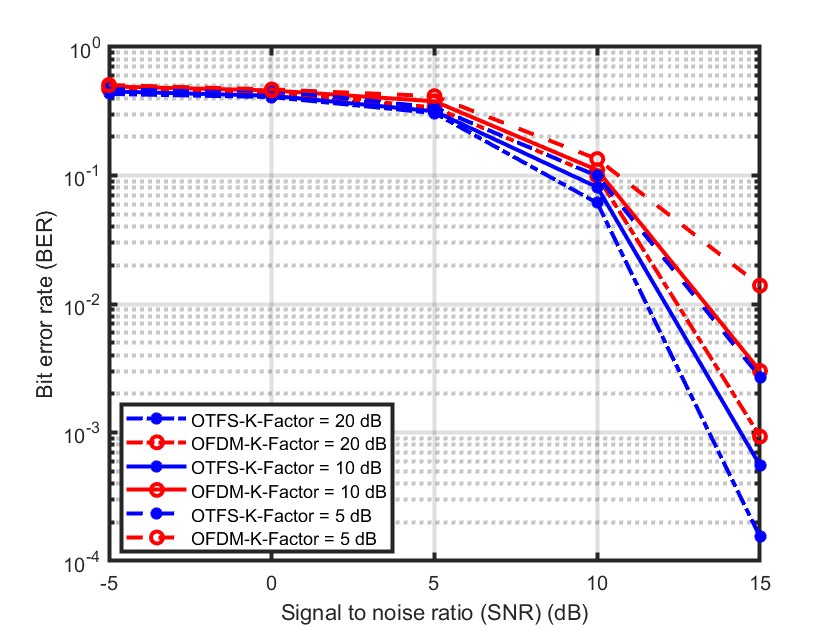}
\centering
    \caption{BER performance comparison between OTFS and OFDM, with varying Rician factor, $\kappa_d$ (Height = 10 km, Speed of HAP = 150 m/s, antenna dimension = $100 \times 100$)}
\label{fig20}
\end{figure}

\begin{figure}[h]
\includegraphics[width=2.8in,height=2.05in]{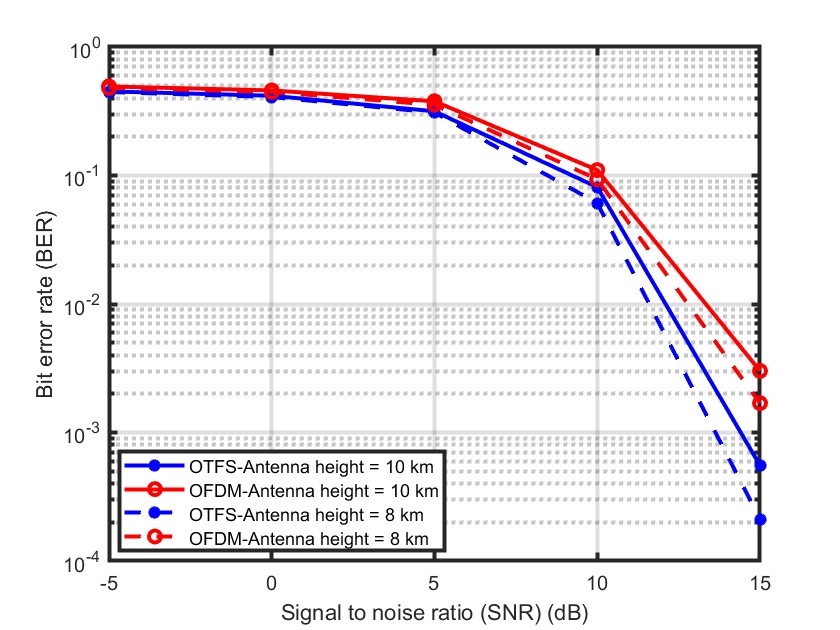}
\centering
    \caption{BER performance comparison between OTFS and OFDM, with varying airliner height ($\kappa_d$ = 10 dB, velocity of HAP = 150 m/s, antenna dimension = $100 \times 100$)}
\label{fig21}
\end{figure}

\begin{figure}[h]
\includegraphics[width=2.8in,height=2.05in]{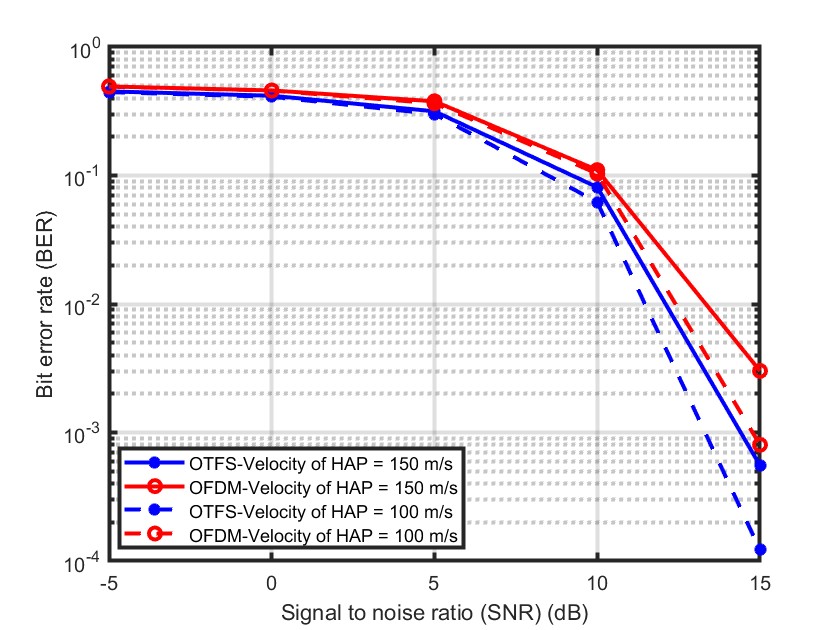}
\centering
    \caption{BER performance comparison between OTFS and OFDM, with varying velocity of HAP (Height= 10 km, $\kappa_d$  = 10 dB, antenna dimension = $100 \times 100$)}
\label{fig22}
\end{figure}

\begin{figure}[h]
\includegraphics[width=2.8in,height=2.05in]{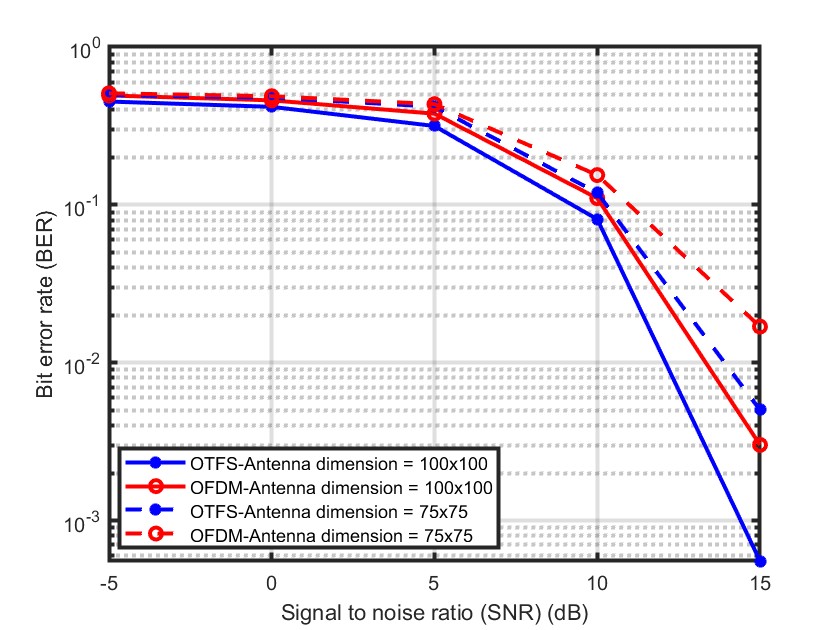}
\centering
    \caption{BER performance comparison between OTFS and OFDM, with varying antenna dimensions (Height= 10 km, velocity of the HAP = 150 m/s, $\kappa_d$ = 10 dB)}
\label{fig23}
\end{figure}

In Figs. \ref{fig20}-\ref{fig23}, we compare the \ac{BER} performance of OTFS and OFDM systems. Fig. \ref{fig20} presents the scenario where an airliner/HAP moves at a speed of $150$ m/s, with system parameters set to $M = 512$, $N = 16$, and an antenna array dimension of $100 \times 100$, for three
different Rician factors, $\kappa_d$. The results clearly indicate that, under high Doppler conditions, OTFS outperforms OFDM, offering superior resilience to time-varying channels. For example, with a Rician factor of $10$ dB, OTFS achieves a gain of $2$ dB over OFDM at a BER of $10^{-2}$. A similar trend is observed at the varying height of the airliner, as shown in Fig. \ref{fig21}. OTFS continues to show lower BER than OFDM at every altitude tested, indicating its robustness even as the propagation distance increases. This further highlights OTFS’s ability to maintain reliable performance under high-mobility scenarios. In Fig. \ref{fig22}, we demonstrate the superiority of OTFS over OFDM for varying velocity of the airliner. With an increase in velocity from $100$ m/s to $150$ m/s, the drop in performance is smaller. OTFS shows remarkable resilience across different speeds of \ac{HAP}. By utilizing \ac{DD} mapping, \ac{OTFS} distinguishes signals despite rapid channel changes, allowing it to maintain consistent performance even at higher speeds.
\begin{table}[!h]
\centering
\caption{Simulation parameters}
\label{tab:allmethods}

\renewcommand{\arraystretch}{0.98} 

  \begin{center}
    \begin{tabular}{|p{5.1cm}|p{1.8cm}|}
    \hline
    \centering
    \textbf{Parameter} & \textbf{Values} \\
\hline 
      Carrier frequency (GHz) &   28  \\
      \hline 
      Macro-cell radius $\textit{R}$ (km) &   8  \\
\hline 
      Micro-cell radius $\textit{r}$ (m) &   75  \\
      \hline 
      Vertical airliner/HAP distance $H_{t}$ (km) &   10 \\
      \hline 
      Velocity of HAP (m/s) &  150 \\
      \hline
Size of OTFS symbol ($M,N$) & (512,16) \\
\hline Subcarrier spacing (kHz) &  30\\

\hline   
Number of \ac{LoS} path &  1  \\
\hline
Number of \ac{NLoS} path &  1  \\
\hline
Delay spread (ns) &  150  \\
\hline
Total Bandwidth (MHz) &  15.36  \\
\hline
Reuse factor &  7  \\
\hline
Dimensions of the uniform planar array  &  $100\times100$ \\
\hline
Maximum range  (km) &  10 \\
\hline
Maximum speed (m/s) &  160.7 \\
\hline
Cyclic Prefix &  64  \\
\hline
Rician factor $\kappa_{d}$ (dB)&  10  \\
\hline
Back-Off (dBm) &  10 \\
\hline
Transmitter antenna gain &  10 log ($L^{2}$) \\
\hline
Transmit power (dBm)& 5\\
\hline
Atmospheric and cloud loss (dB) & 7.9 \\
\hline
Receiver antenna gain (dB) & 60.2 \\
\hline
Receiver noise figure (dB) & 6 \\
\hline
Other receiver loss (dB) & 1.8\\
\hline
\end{tabular}
\vspace{-2mm}
\end{center}
\end{table}

Fig. \ref{fig23} compares the performance of \ac{OTFS} and \ac{OFDM} modulation schemes for different antenna dimensions ($100\times100$ and $75\times75$). The larger antenna array ($100\times100$) provides a greater beamforming gain, which enhances the signal strength at the receiver. With an array of $75\times75$ antennas, there are fewer antenna elements, leading to a reduced array gain. The resulting weaker signal amplification makes noise and interference more dominant, which in turn raises the \ac{BER}. In both configurations ($100\times100$ and $75\times75$), OTFS consistently outperforms OFDM, achieving a lower BER at the same SNR. For example, with an antenna configuration of $75\times75$, OTFS achieves a gain of $2$ dB over OFDM at a BER of $10^{-2}$. The results in Figs. \ref{fig20}-\ref{fig23} clearly show that OTFS achieves superior BER performance compared to OFDM in all the conditions evaluated. This performance gain remains consistent with changes in $\kappa_d$, velocity, height, and antenna dimensions, highlighting the robustness of OTFS under various propagation and system conditions. 
Thus, our proposed framework provides insight into how OTFS can sustain reliable links
in airborne scenarios with large-scale antenna arrays, which is timely and novel compared to existing OTFS studies focused only on ground-based
mobility.

\section{Conclusions}
\label{conclusion}
In this paper, we examine the performance of airplane-aided integrated communication systems in high mobility scenarios, using \ac{OTFS} modulation with a uniform planar antenna array. We also used \ac{NSB} beamforming technique at the transmitter to mitigate multi-tier interference. Experiments are carried out under different settings of airliner height, \ac{HAP} velocity, antenna array dimensions, and Rician factor to analyze various aerial networking conditions. The simulation results clearly show that \ac{OTFS} consistently outperforms conventional \ac{OFDM} in terms of \ac{BER} performance under all tested conditions, confirming its suitability for next-generation aerial communication systems. Using DD-domain processing, OTFS provided improved resilience against Doppler shifts and time-selective channels, ensuring reliable and consistent communication for aerial-terrestrial applications, even in challenging high-mobility environments.

\bibliographystyle{IEEEtran}
\bibliography{ref}

\end{document}